\documentclass[prd,superscriptaddress,amsfonts,amssymb,amsmath,showpacs,twocolumn]{revtex4-2}
\usepackage{bm}
\usepackage{amsfonts}
\usepackage{latexsym}
\usepackage[latin1]{inputenc}
\usepackage{graphicx}
\usepackage{amsmath}
\usepackage{palatino}
\usepackage{mathpazo}
\usepackage{textcomp}
\usepackage{float}
\usepackage{booktabs}
\usepackage{dcolumn}
\usepackage{booktabs}
\usepackage{multirow}
\usepackage{hyperref}
\hypersetup{colorlinks,citecolor=blue}
\usepackage{amsmath}
\usepackage{xcolor}
\usepackage{orcidlink}
\usepackage[caption=false]{subfig}
\usepackage{commath}
\def\jnl@style{\it}
\def\aaref@jnl#1{{\jnl@style#1}}

\def\aaref@jnl#1{{\jnl@style#1}}

\def\aj{\aaref@jnl{AJ}}                   % Astronomical Journal
\def\apj{\aaref@jnl{ApJ}}                 % Astrophysical Journal
\def\apjl{\aaref@jnl{ApJ}}                % Astrophysical Journal, Letters
\def\apjs{\aaref@jnl{ApJS}}               % Astrophysical Journal, Supplement
\def\apss{\aaref@jnl{Ap\&SS}}             % Astrophysics and Space Science
\def\aap{\aaref@jnl{A\&A}}                % Astronomy and Astrophysics
\def\aapr{\aaref@jnl{A\&A~Rev.}}          % Astronomy and Astrophysics Reviews
\def\aaps{\aaref@jnl{A\&AS}}              % Astronomy and Astrophysics, Supplement
\def\mnras{\aaref@jnl{Mon.~Not.~Roy.~Astron.~Soc.}}             % Monthly Notices of the RAS
\def\prd{\aaref@jnl{Phys.~Rev.~D}}        % Physical Review D
\def\prc{\aaref@jnl{Phys.~Rev.~C}}  % Physical Review C
\def\prl{\aaref@jnl{Phys.~Rev.~Lett.}}    % Physical Review Letters
\def\qjras{\aaref@jnl{QJRAS}}             % Quarterly Journal of the RAS
\def\skytel{\aaref@jnl{S\&T}}             % Sky and Telescope
\def\ssr{\aaref@jnl{Space~Sci.~Rev.}}     % Space Science Reviews
\def\zap{\aaref@jnl{ZAp}}                 % Zeitschrift fuer Astrophysik
\def\nat{\aaref@jnl{Nature}}              % Nature
\def\aplett{\aaref@jnl{Astrophys.~Lett.}} % Astrophysics Letters
\def\apspr{\aaref@jnl{Astrophys.~Space~Phys.~Res.}} % Astrophysics Space Physics Research
\def\physrep{\aaref@jnl{Phys.~Rep.}}      % Physics Reports
\def\physscr{\aaref@jnl{Phys.~Scr}}       % Physica Scripta
\def\commat{\aaref@jnl{Comm.~Math.~Phys.}}              % Communications in Mathematical Physics
\def\science{\aaref@jnl{Science}}               % Science
\def\cqg{\aaref@jnl{Classical Quant.~Grav.}}            % Classical and Quantum Gravity
\def\jpcs{\aaref@jnl{JPCS}}                                     % Journal of Physics Conference Series
\def\ijmpd{\aaref@jnl{Int.~J.~Mod.~Phys.~D}}                    % International Journal of Modern Physics D
\def\grg{\aaref@jnl{Gen.~Relat.~Gravit.}}               % General Relativity and Gravitation
\def\rpp{\aaref@jnl{Rep.~Prog.~Phys.}}          % Reports on Progress in Physics
\def\npa{\aaref@jnl{Nucl.~Phys.~A}}        % Nuclear Physics A
\def\lrr{\aaref@jnl{Living Rev.~Rel.}}                   % Living reviews in relativity
\def\jcap{\aaref@jnl{J.~Cosmology Astropart.~Phys.}}    % Journal of cosmology and astroparticle physics
\def\rmp{\aaref@jnl{Rev.~Mod.~Phys.}}   %Reviews of modern physics
\def\epjc{\aaref@jnl{Eur.~Phys.~J.~C}}

\allowdisplaybreaks[1]
\begin{document}

\color{black}       %% For one column

\title{Quintessence Universe and cosmic acceleration in $f(Q,T)$ gravity}
%\end{document}
\author{M. Koussour\orcidlink{0000-0002-4188-0572}}
\email{pr.mouhssine@gmail.com}
\affiliation{Quantum Physics and Magnetism Team, LPMC, Faculty of Science Ben
M'sik,\\
Casablanca Hassan II University,
Morocco.}

\author{N. Myrzakulov\orcidlink{0000-0001-8691-9939}}
\email{nmyrzakulov@gmail.com}
\affiliation{L. N. Gumilyov Eurasian National University, Nur-Sultan 010008,
Kazakhstan.}
\affiliation{Ratbay Myrzakulov Eurasian International Centre for Theoretical
Physics, Nur-Sultan 010009, Kazakhstan.}

\author{S.H. Shekh\orcidlink{0000-0003-4545-1975}}
\email{da\_salim@rediff.com}
\affiliation{Department of Mathematics. S. P. M. Science and Gilani Arts Commerce
College, Ghatanji, Dist. Yavatmal, Maharashtra-445301, India.}

\author{M. Bennai\orcidlink{0000-0002-7364-5171}}
\email{mdbennai@yahoo.fr }
\affiliation{Quantum Physics and Magnetism Team, LPMC, Faculty of Science Ben
M'sik,\\
Casablanca Hassan II University,
Morocco.} 
\affiliation{Lab of High Energy Physics, Modeling and Simulations, Faculty of
Science,\\
University Mohammed V-Agdal, Rabat, Morocco.}
%{Universidade Federal do ABC (UFABC) - Centro de Ci\^encias Naturais e Humanas (CCNH) - Avenida dos Estados 5001, 09210-580, Santo Andr\’e, SP, Brazil}
%\affiliation{Universidade de S\~ao Paulo (USP), Instituto de Astronomia, Geof\'isica e Ci\^encias Atmosf\'ericas (IAG), Rua do Mat\~ao 1226, Cidade Universit\'aria, 05508-090 S\~ao Paulo, SP, Brazil}
%\affiliation{Instituto Tecnol\'ogico de Aeron\'autica (ITA), Departamento de F\'isica, Centro T\'ecnico Aeroespacial, 12228-900 S\~ao Jos\'e dos Campos, S\~ao Paulo, Brazil}

%%%%%%%%%%%%%%%%%%%%%%%%%%%%%%%%%%%%  DATE  %%%%%%%%%%%%%%%%%%%%%%%%%%%%%%%%%%%%
\date{\today}

\begin{abstract}
The problem of cosmic acceleration and dark energy is one of the mysteries
presently posed in the scientific society that general relativity has not
been able to solve. In this work, we have considered alternative models to
explain this late-time acceleration in a flat Friedmann-Lemaitre-Robertson-Walker (FLRW) Universe within the framework of the $%
f\left(Q,T\right) $\ modified gravity theory (where $Q$ is the non-metricity
and $T$ is the trace of the energy-momentum tensor) recently proposed by Xu
et al. (Eur. Phys. J. C 79 (2019) 708), which is an extension of $f\left(
Q\right) $ gravity with the addition of the $T$ term. Here, we presume a
specific form of $f\left( Q,T\right) =\alpha Q+\beta Q^{2}+\gamma T$ where $%
\alpha $, $\beta $ and $\gamma $ are free model parameters, and obtained the
exact solutions by assuming the cosmic time-redshift relation as $t(z)=\frac{%
nt_{0}}{m}g(z)$ which produces the Hubble parameter of the form $H(z)=\frac{%
mH_{0}}{m+n}\left[ \frac{1}{g(z)}+1\right] $ where $m$ and $n$ are the
non-negative constants, we find the best values for them using 57 data
points of the Hubble parameter $H\left( z\right) $. Also, we find the
behavior of different cosmological parameters as the deceleration parameter $%
\left( q\right) $, energy density $\left( \rho \right) $, pressure $%
\left(p\right) $ and EoS parameter $\left( \omega \right) $ and compare them
with the observational results. To ensure the validity of the results, we
studied the energy conditions along with jerk parameter. Finally, we found
that our model behaves similarly to the quintessence Universe.
\end{abstract}
\maketitle

\section{Introduction} \label{sec1}
General Relativity (GR) is based that space and time constitute a unified
structure assigned on Riemannian manifolds with the metric and the Levi-Civita
connection. As we know GR is established on some main assumptions like 
\textit{Relativity principle}, \textit{Equivalence principle}, \textit{%
General Covariance principle}, \textit{Causality principle} and \textit{%
Lorentz covariance}. Moreover, it is well-known that GR has two other
equivalent descriptions, that are based on different connections. When we
investigate a metric compatible but flat connection (i.e. the curvature is
zero), we get the Teleparallel Equivalent of General Relativity (TEGR),
where gravity is formulated through torsion \cite{ref1}. Finally, the last
description is Symmetric Teleparallel Equivalent of GR (STEGR) which works
in a flat and torsion-free connection where non-metricity is assumed as
gravitational interactions \cite{ref2} . On the other hand, modern observations in cosmology of SNIa (type Ia
Supernova) \cite{SN1, SN2}, LSS (Large Scale Structure) \cite{LS1, LS2},
WMAP (Wilkinson Microwave Anisotropy Probe) data \cite{WMAP1, WMAP2, WMAP9},
CMB (Cosmic Microwave Background) \cite{CMB1, CMB2}, and BAO (Baryonic
Acoustic Oscillations) \cite{BAO1, BAO2} show that the expansion of the
Universe has entered an acceleration phase. Moreover, the same observational
data display that everything we see around us is only $5\%$ of the total
content of the Universe, and the remaining content, i.e. $95\%$ is in the
form of unknown species dubbed Dark Matter (DM) and Dark Energy (DE). The
results of these observations contradict GR, in particular the standard
Friedmann equations, which are part of the applications of GR on a
homogeneous and isotropic Universe on a large scale. Consequently, GR is not
the final theory of gravity, it might be a special case of a more general
theory of gravity.

To account for the recent observational data associated with the acceleration of the Universe, in recent years in the literature, there is a
huge effort to modify gravity in order to be able to describe the evolution
of the Universe and solve the mysteries of DE and DM. Most of the works
start from the curvature-based Einstein-Hilbert action formulation and
extend it in the form $f(R)$ \cite{ref3}. Also successfully built a
gravitational modification torsion-based on TEGR namely $f(T)$ gravity \cite%
{ref4,ref5}. Note that TEGR at the level of equations coincides completely
with general relativity, but equations of their modifications are different
because field equations of $f(T)$ gravity are of second order while those of 
$f(R)$ are of fourth-order. Very recently new modified $f(Q)$ gravity theory
was proposed as a geometric interpretation and attracted a lot of attention
in which gravity is attributed to non-metricity, which geometrically
describes the variation of the length of a vector in the parallel transport 
i.e. $Q_{\gamma \mu \nu }=\nabla _{\gamma }g_{\mu \nu }$ \cite{ref6,ref7}.
The differential geometry in this case is called Weyl geometry which is a
generalization of Riemannian geometry i.e. the geometric basis of GR. These
non-metricity-based $f(Q)$ theories represent a generalization of the STEGR,
like $f(R)$ and $f(T)$ gravity. The theory $f(Q)$ gravity has been explored
in different contexts and cosmological applications. In \cite{ref8}
investigated the evolution of linear perturbations in the $f(Q)$ gravity and
considered different evolutions of the effective dark energy equation of
state. Non-metricity scalar $Q$ and the equations of motion for generic
static and spherically symmetric geometry with an anisotropic fluid is
derived in \cite{ref9}. Application of Diracs method for the quantization of
constrained systems in the context of $f(Q)$ gravity is presented in \cite%
{ref10}. In \cite{ref11, Koussour1} derived the gravitational equations for $%
f(Q)$ gravity in the homogeneous, anisotropic locally, rotationally,
symmetric Bianchi-I Universe in the presence of a single anisotropic perfect
fluid.

One of interesting extensions of symmetric teleparallel gravity newly proposed as geometric alternatives to DE based on the coupling between non-metricity $Q$ and the trace of the energy-momentum tensor $T$, i.e., considering an arbitrary function $f(Q,T)$ in the gravitational action \cite%
{ref12}. It is clear that for $T=0$ i.e. the case of vacuum, this theory
reduces to the $f(Q)$ gravity, which is equivalent to GR and passes all
solar system tests. The full set of field equations of this theory are
obtained by varying the gravitational action with respect to both metric and
connection, separately. The covariant divergence of the gravitational
equations are obtained. Such coupling can lead to the non-conservation of
the energy-momentum tensor. This conservation violation has substantial
physical clues that predict large changes in the thermodynamics of the
Universe similar to those predicted by $f(R,T)$ gravity. Note that the
resulting theory differs from well-known $f(R,T)$ gravity \cite{ref13} and $%
f(T,\mathcal{T})$ gravity \cite{ref14} in that it is a novel modified
gravitational theory based on a more general geometric framework than
Riemannian geometry (Weyl geometry), with no curvature-equivalent and no
torsion-equivalent, and its cosmological implications are very interesting.
In the literature, there are active works in the framework $f(Q,T)$ gravity
theory \cite{ref15,ref16, ref17,ref18,ref19}. Thus, there is a motivation to
examine several theoretical, observational and cosmological aspects of $%
f(Q,T)$ gravity. It has newly been found out that $f(Q,T)$ gravity
dramatically alters the nature of tidal forces and the equation of motion in
the Newtonian limit \cite{Yang/2021}. In \cite{ref20} authors developed the
cosmological linear theory of perturbations for $f(Q,T)$ gravity. They claim
that results might also enable to test with CMB and standard siren data.
Energy conditions constraints on different forms of $f(Q,T)$ gravity was
investigated in \cite{ref21,ref22}. This analysis were carried using the
actual values of the deceleration parameters, Hubble and verify the
compatibility with $\Lambda $CDM model. Weyl form of $f(Q,T)$ gravity model
in which the scalar non-metricity is fully determined by a vector field $%
w_{\mu }$ proposed in \cite{ref23}. This gravity theory can be considered as
an useful and alternative approach for the description of the early phases
and late phases of cosmological evolution.

In this work, we have also investigated the cosmological model with
Friedmann-Lemaitre-Robertson-Walker (FLRW) Universe in $f(Q,T)$ theory
taking into account the coupling of the form $f\left( Q,T\right) =\alpha
Q+\beta Q^{2}+\gamma T$, where $\alpha $, $\beta $, and $\gamma $ are free
model parameters. This supported by $f\left( R,T\right) $ gravity form $%
f\left( R,T\right) =\alpha R+\beta R^{2}+\gamma T$ in which the presence of
square term of $R$ reveals the existence of the late-time acceleration
phase. Using the hybrid expansion law of the scale factor which leads to the
time-redshift relation in the form of the Lambert function i.e. $t(z)=\frac{%
nt_{0}}{m}g(z)$ that has been studied in several modified theories of
gravity \cite{Koussour2, Koussour3}, we have analyzed the various
cosmological parameters like deceleration parameter, energy density,
pressure, and the equation of state (EoS) parameter with the energy
conditions for our cosmological model. In addition, we try to constrain the
model parameters using the recent $57$ Hubble datasets points by minimizing
the chi-square function.

The manuscript is organizing in the following form: In Sect. \ref{sec2}
presented field equations of the theory by varying action.
Gravitational field equations along with it's solution for the FLRW line
element are shown in Sect. \ref{sec3}. Comparison with observational data
constraints from $H(z)$ datasets demonstrated in Sect. \ref{sec4}. In Sect. %
\ref{sec5}, we studied some physical parameters including energy conditions
and the Cosmographic jerk parameter for particular case of $f(Q,T)$ theory while the conclusions is given in the last section in detail.

\section{Basic formalism in $f\left( Q,T\right) $ gravity}

\label{sec2}

The modified Einstein-Hilbert action for the $f(Q,T)$ extended symmetric
teleparallel gravity is given by \cite{ref12}

\begin{equation}
S=\int \left( \frac{1}{16\pi }f(Q,T)+L_{m}\right) \sqrt{-g}d^{4}x,
\label{eqn1}
\end{equation}%
where $f(Q,T)$ being the general function of the non-metricity scalar $Q$
and the trace of the energy-momentum tensor $T$, $g$ is the determinant of
the metric tensor $g_{\mu \nu }$ i.e. $g=\det \left( g_{\mu \nu }\right) $,\
and $L_{m}$ is the usual matter Lagrangian.\ The non-metricity scalar $Q$ is
defined as

\begin{equation}
Q\equiv -g^{\mu \nu }(L_{\,\,\,\alpha \mu }^{\beta }L_{\,\,\,\nu \beta
}^{\alpha }-L_{\,\,\,\alpha \beta }^{\beta }L_{\,\,\,\mu \nu }^{\alpha }),
\label{eqn2}
\end{equation}%
where the disformation tensor $L^{\delta }{}_{\alpha \gamma }$ is given by 
\begin{equation}
L_{\alpha \gamma }^{\beta }=-\frac{1}{2}g^{\beta \eta }(\nabla _{\gamma
}g_{\alpha \eta }+\nabla _{\alpha }g_{\eta \gamma }-\nabla _{\eta }g_{\alpha
\gamma }).  \label{eqn3}
\end{equation}

The non-metricity tensor is defined by the following form 
\begin{equation}
Q_{\gamma \mu \nu }=\nabla _{\gamma }g_{\mu \nu },  \label{eqn4}
\end{equation}%
and the trace of the non-metricity tensor is obtained as 
\begin{equation}
Q_{\beta }=g^{\mu \nu }Q_{\beta \mu \nu }\qquad \widetilde{Q}_{\beta
}=g^{\mu \nu }Q_{\mu \beta \nu }.  \label{eqn5}
\end{equation}

Further, we define the superpotential tensor as follows 
\begin{equation}
P_{\,\,\,\mu \nu }^{\beta }=-\frac{1}{2}L_{\,\,\,\mu \nu }^{\beta }+\frac{1}{%
4}(Q^{\beta }-\widetilde{Q}^{\beta })g_{\mu \nu }-\frac{1}{4}\delta _{(\mu
}^{\beta }Q_{\nu )},  \label{eqn6}
\end{equation}%
and using this definition above, the non-metricity scalar is given as 
\begin{equation}
Q=-Q_{\beta \mu \nu }P^{\beta \mu \nu }.  \label{eqn7}
\end{equation}

Here, the definition of the energy-momentum tensor of the matter is given by 
\begin{equation}
T_{\mu \nu }=-\frac{2}{\sqrt{-g}}\dfrac{\delta (\sqrt{-g}L_{m})}{\delta
g^{\mu \nu }},  \label{eqn8}
\end{equation}%
and 
\begin{equation}
\Theta _{\mu \nu }=g^{\alpha \beta }\frac{\delta T_{\alpha \beta }}{\delta
g^{\mu \nu }}.  \label{eqn9}
\end{equation}

The variation of energy-momentum tensor with respect to the metric tensor $%
g_{\mu \nu }$\ read as 
\begin{equation}
\frac{\delta \,g^{\,\mu \nu }\,T_{\,\mu \nu }}{\delta \,g^{\,\alpha \,\beta }%
}=T_{\,\alpha \beta }+\Theta _{\,\alpha \,\beta }.  \label{eqn10}
\end{equation}

In addition, the fleld equations of $f\left( Q,T\right) $ gravity are given
by varying the action $\left( S\right) $ with respect to metric tensor $%
g_{\mu \nu }$ 
\begin{widetext}
\begin{equation}
-\frac{2}{\sqrt{-g}}\nabla _{\beta }(f_{Q}\sqrt{-g}P_{\,\,\,\,\mu \nu
}^{\beta }-\frac{1}{2}fg_{\mu \nu }+f_{T}(T_{\mu \nu }+\Theta _{\mu \nu
})-f_{Q}(P_{\mu \beta \alpha }Q_{\nu }^{\,\,\,\beta \alpha }-2Q_{\,\,\,\mu
}^{\beta \alpha }P_{\beta \alpha \nu })=8\pi T_{\mu \nu },  \label{eqn11}
\end{equation}%
\end{widetext}
where $f_{Q}=\frac{df\left( Q,T\right) }{dQ}$, $f_{T}=\frac{df\left(
Q,T\right) }{dT}$, and $\nabla _{\beta }$\ denotes the covariant derivative.
From Eq. (\ref{eqn11}) it appears that the field equations of $f\left(
Q,T\right) $ extended symmetric teleparallel gravity depends on the tensor $%
\theta _{\mu \nu }$. Therefore, depending on the nature of the source of
matter, different cosmological models of $f\left( Q,T\right) $ gravity are
possible. Originally, Xu et al. \cite{ref12} obtained three models using
following functional forms of $f\left( Q,T\right) $: (i) $f\left( Q,T\right)
=\alpha Q+\beta T$, (ii) $f\left( Q,T\right) =\alpha Q^{n+1}+\beta T$, (iii) 
$f\left( Q,T\right) =-\alpha Q-\beta T^{2}$ where $\alpha $, $\beta $ and $n$
are constants.

\section{Flat FLRW Universe in $f\left( Q,T\right) $ cosmology}

\label{sec3}

To solve field equations in $f(Q,T)$ extended symmetric teleparallel gravity
it is usually necessary to make simplifying assumptions such as the choice
of a metric. Therefore, in this work, we consider the spatially homogeneous
and isotropic flat FLRW Universe given by the following metric, 
\begin{equation}
ds^{2}=-N^{2}\left( t\right) dt^{2}+a^{2}(t)\left(
dx^{2}+dy^{2}+dz^{2}\right) ,  \label{eqn12}
\end{equation}%
where $a\left( t\right) $ is the scale factor of the Universe, depending
only on cosmic time (where the cosmic time is measure in Gyr) and $N\left(
t\right) $\ is the lapse function considered to be $1$ in the standard case.
The rates of expansion and dilation are determined as $H\equiv \frac{\overset%
{.}{a}}{a}$, $T\equiv \frac{\overset{.}{N}}{N}$ respectively. Thus, the
corresponding non-metricity scalar is given by $Q=6\frac{H^{2}}{N^{2}}$. In
this article, we presume that the content of the Universe as a perfect fluid
whose energy-momentum tensor is given by 
\begin{equation}
T_{\nu }^{\mu }=diag\left( -\rho ,p,p,p\right) ,  \label{eqn13}
\end{equation}

Here, $p$ is the perfect fluid pressure and $\rho $ is the energy density of
the Universe. Thus, for tensor $\theta _{\nu }^{\mu }$ the expression is
obtained as $\theta _{\nu }^{\mu }=diag\left( 2\rho +p,-p,-p,-p\right) $.
Considering the case as $N=1$, the Einstein field equations using the metric
(\ref{eqn12}) are given as, 
\begin{equation}
8\pi \rho =\frac{f}{2}-6FH^{2}-\frac{2\widetilde{G}}{1+\widetilde{G}}\left( 
\overset{.}{F}H+F\overset{.}{H}\right) ,  \label{eqn14}
\end{equation}%
\begin{equation}
8\pi p=-\frac{f}{2}+6FH^{2}+2\left( \overset{.}{F}H+F\overset{.}{H}\right) .
\label{eqn15}
\end{equation}%
where, we used $Q=6H^{2}$ and $\left( \text{\textperiodcentered }\right) $
dot represents a derivative with respect to cosmic time $\left( t\right) $.
In this case, $F\equiv f_{Q}$ and $8\pi \widetilde{G}\equiv f_{T}$ represent
differentiation with respect to $Q$ and $T$ respectively. The evolution
equation for the Hubble function $H$ can be obtained by combining Eqs. (\ref%
{eqn14}) and (\ref{eqn15}) as, 
\begin{equation}
\overset{.}{H}+\frac{\overset{.}{F}}{F}H=\frac{4\pi }{F}\left( 1+\widetilde{G%
}\right) \left( \rho +p\right) .  \label{eqn16}
\end{equation}

Einstein's field equations (\ref{eqn14}) and (\ref{eqn15}) can be construed
as extended symmetric teleparallel equivalents to Friedman's equations with
additional terms from the non-metricity of space-time and the trace of the
energy-momentum tensor $T$ which behaves as an effective component. Thus,
the effective energy density $\rho _{eff}$\ and effective pressure $p_{eff}$%
\ are defined by 
\begin{widetext}
\begin{equation}
3H^{2}=8\pi \rho _{eff}=\frac{f}{4F}-\frac{4\pi }{F}\left[ \left( 1+%
\widetilde{G}\right) \rho +\widetilde{G}p\right] ,  \label{eqn17}
\end{equation}%
\begin{equation}
2\overset{.}{H}+3H^{2}=-8\pi p_{eff}=\frac{f}{4F}-\frac{2\overset{.}{F}H}{F}+%
\frac{4\pi }{F}\left[ \left( 1+\widetilde{G}\right) \rho +\left( 2+%
\widetilde{G}\right) p\right] .  \label{eqn18}
\end{equation}
\end{widetext}

In order to obtain the exact solutions to the field equations above, we
assume the Lambert function distribution for the time-redshift relation $%
t\left( z\right) $ as following%
\begin{equation}
t\left( z\right) =\frac{nt_{0}}{m}g\left( z\right) ,  \label{eqn19}
\end{equation}%
and%
\begin{equation}
g\left( z\right) =LambertW\left[ k\left( 1+z\right) ^{-\frac{1}{n}}\right] ,
\label{eqn20}
\end{equation}%
where $k=\frac{m}{n}e^{\frac{m}{n}}$, $n$ and $m$ are non-negative constants
and $t_{0}$ represents the present age of the Universe. Using the relation
between the scale factor and redshift of the Universe $a\left( t\right)
=a_{0}\left( 1+z\right) ^{-1}$ where $a_{0}$ represent the present value of
scale factor, we find the Hubble parameter as%
\begin{equation}
H=\frac{\overset{.}{a}}{a}=\frac{m}{t_{0}}+\frac{n}{t}.  \label{eqn21}
\end{equation}

The motivation behind the above choice is that the relation (\ref{eqn21})
produces the scale factor of the hybrid type, and it is known in the
literature that this type depicts the transition from the early decelerating
phase to the present accelerating phase, which create a time-dependent
deceleration parameter. Also, the ansatz reduces to the usual power law for $%
m=0$ and de Sitter solutions for certain values of $m$ and $n$. The
corresponding deceleration parameter is given as 
\begin{equation}
q=-1+\frac{d}{dt}\left( \frac{1}{H}\right) =-1+nt_{0}^{2}\left(
nt_{0}+mt\right) ^{-2}.  \label{eqn22}
\end{equation}

\begin{figure}[H]
\centering
\includegraphics[scale=0.67]{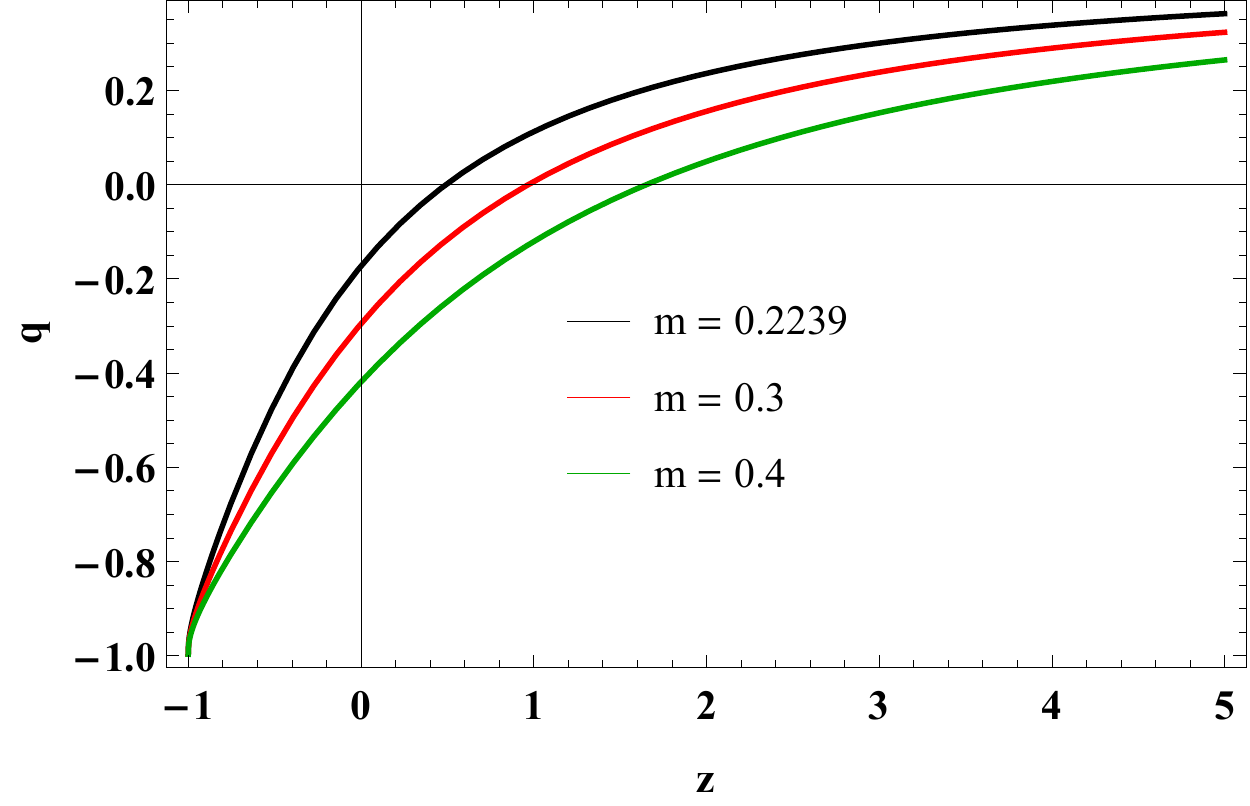}
\caption{Deceleration parameter $\left( q\right) $ versus redshift $\left(
z\right) $ with $n=0.6886$.}
\label{fig1}
\end{figure}

For the model constant parameters $m$ and $n$, we found the appropriate
values in the next section using the observational Hubble data $H\left(
z\right) $ (OHD) as $m=0.2239$ and $n=0.6886$. Fig. \ref{fig1} clearly shows
the transition of the Universe from the deceleration phase $\left(
q>0\right) $ to the acceleration phase $\left( q<0\right) $ with the
transition redshift $z_{tr}=0.5234$ for $m=0.2239$. Thus, the transition
redshift value for our model is in conformity with the observational data.

In addition, in Fig. \ref{fig1} we examine the effect of parameter $m$ on
the model through three different values, namely $m=0.2239,0.3,0.4$.

\section{Observational constraints from $H(z)$ datasets}

\label{sec4}

To get the best fit value of the model parameters $m$ and $n$ of our model
under study and to compare our results with observation data, we need to
constrain the parameters using some observational datasets. In this section,
we will use Hubble datasets with $57$ data points. In \cite{ref24} Sharov
and Vasiliev prepared a list of $57$ data points of measurements of the
Hubble parameter in the cosmological redshift range $0.07\leq z\leq 2.42$, $%
31$ points from the differential age method (DA method)\ and the other $26$
points were evaluated using BAO data and other methods (See Tab. \ref{tab1}).

\begin{figure}[H]
\centering
\includegraphics[scale=0.39]{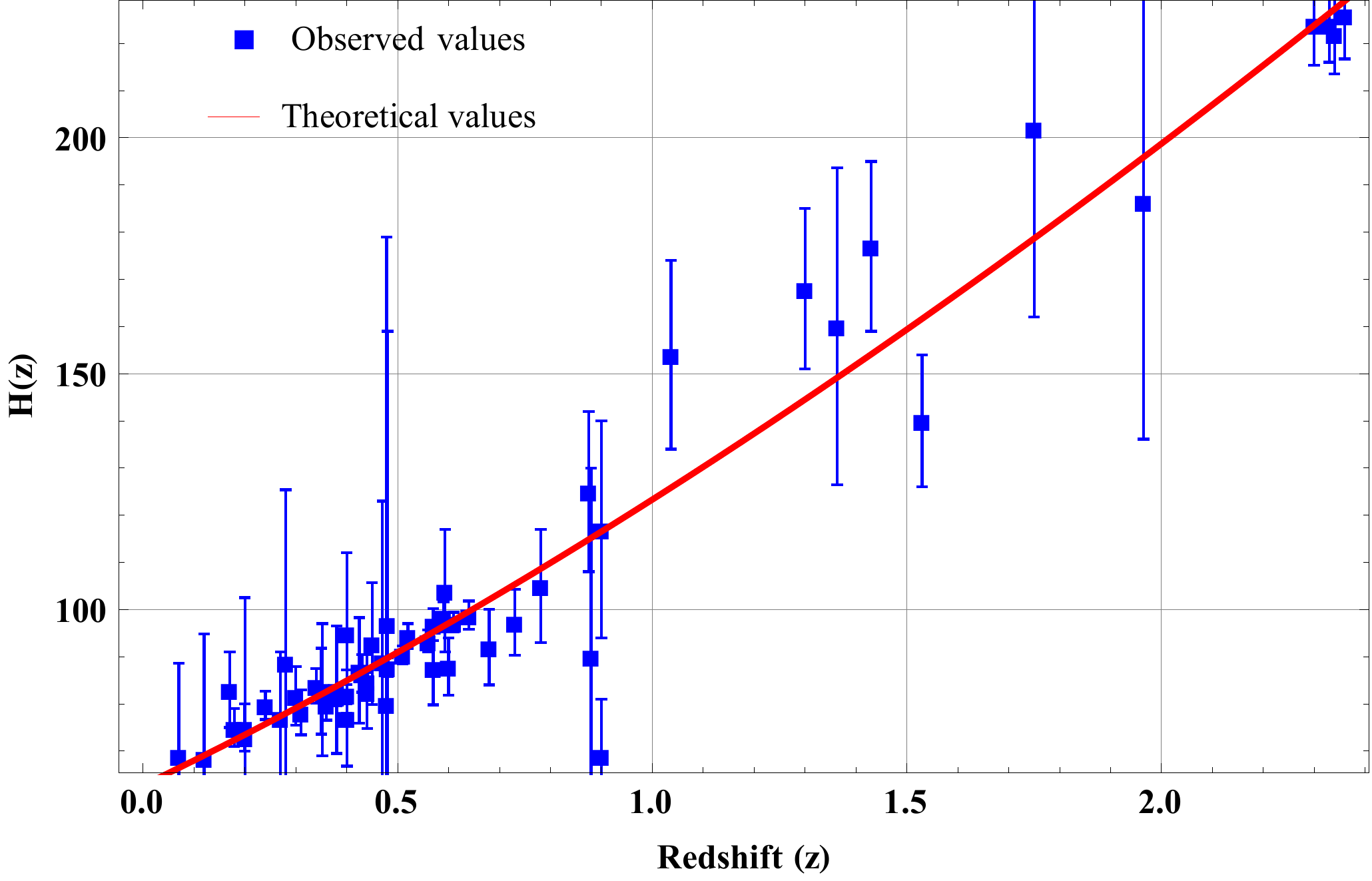}
\caption{Best fit curve of Hubble function $H\left( z\right) $ versus
redshift $z$. The blue dots represents error bars of 57 data points, the red
line is the curve gained for our model.}
\label{fig2}
\end{figure}

Using Eqs. (\ref{eqn19})-(\ref{eqn21}), the Hubble parameter $H$ in terms of
the cosmological redshift $z$ as 
\begin{equation}
H(z)=\frac{mH_{0}}{m+n}\left[ \frac{1}{g\left( z\right) }+1\right] ,
\label{eqn23}
\end{equation}%
where $H_{0}=\frac{m+n}{t_{0}}$ is the present value of Hubble parameter.
From Eq. (\ref{eqn23}) we can see that the parameters of the model which are
needed to constrain are $m,n$ and $H_{0}$. Thus, the best fit values of
model parameters $m,n$ and $H_{0}$ are determined by minimizing the
following chi-square function 
\begin{equation}
\chi _{OHD}^{2}(m,n,H_{0})=\sum_{i=1}^{57}\frac{\left[
H_{th}(m,n,H_{0},z_{i})-H_{obs}(z_{i})\right] ^{2}}{\sigma _{i}^{2}},
\label{eqn24}
\end{equation}%
where $H_{th}(m,n,H_{0},z_{i})$ and $H_{obs}(z_{i})$ are theoretical and
observed values of Hubble parameter respectively, and $\sigma _{i}^{2}$
represents the standard error in the observed Hubble parameter measurements.
Also, $\sigma _{i}^{2}$ errors of differential age method and BAO and other
methods are represented in Tab. \ref{tab1}. The best fit values of model
parameters $m,n$ and $H_{0}$ is obtained as $m=0.2239$ ($-0.0962$, $0.544$), 
$n=0.6886$ ($0.4018$, $0.9754$) and $H_{0}=62.73km.s^{-1}.Mpc^{-1}$\ ($54.3$%
, $71.16$) respectively. Fig. \ref{fig2} shows the best fit curve of $%
H\left( z\right) $ versus the cosmological redshift $z$ using 57 Hubble
parameter measurements.$.$

\section{Cosmological $f\left( Q,T\right) $ model}

\label{sec5}

Many researchers have studied several models of $f\left( R,T\right) $
gravity in the form $f\left( R,T\right) =\alpha R+\beta R^{2}+\gamma T$
(where $R$ is the Ricci scala and $T$ is the trace of energy-momentum
tensor) and obtained good results, with the factor $R^{2}$\ is added to
explain the late time acceleration in the expansion of the Universe.
Motivated by this research and by replacing $R$ by $Q$, we obtain 
\begin{equation}
f\left( Q,T\right) =\alpha Q+\beta Q^{2}+\gamma T  \label{eqn25}
\end{equation}%
where $\alpha $, $\beta $ and $\gamma $ are model parameters. The values of $%
f_{Q}$ and $f_{T}$ in the field equations (\ref{eqn14}) and (\ref{eqn15})
are derived as $F=f_{Q}=\alpha +2\beta Q$ and $8\pi \widetilde{G}%
=f_{T}=\gamma $. Using Eq. (\ref{eqn25}) in Eqs. (\ref{eqn14}) and (\ref%
{eqn15} the values of the energy density $\rho $, pressure $p$ and equation
of state (EoS)\ parameter $\omega =\frac{p}{\rho }$ are obtained as

\begin{equation}
\rho =\frac{\alpha \gamma \overset{.}{H}-3H^{2}(8\pi \alpha +\gamma (\alpha
-12\beta \overset{.}{H}))-54\beta (\gamma +8\pi )H^{4}}{2(\gamma +4\pi
)(\gamma +8\pi )}, 
\label{eqn26}
\end{equation}
\begin{equation}
p=\frac{(3\gamma +16\pi )\overset{.}{H}\left( \alpha +36\beta H^{2}\right)
+3(\gamma +8\pi )H^{2}\left( \alpha +18\beta H^{2}\right) }{2(\gamma +4\pi
)(\gamma +8\pi )},  \label{eqn27}
\end{equation}
\begin{equation}
\omega =\frac{(3\gamma +16\pi )\overset{.}{H}\left( \alpha +36\beta
H^{2}\right) +3(\gamma +8\pi )H^{2}\left( \alpha +18\beta H^{2}\right) }{%
\alpha \gamma \overset{.}{H}-3H^{2}(8\pi \alpha +\gamma (\alpha -12\beta 
\overset{.}{H}))-54\beta (\gamma +8\pi )H^{4}}.  \label{eqn28}
\end{equation}

\subsection{Evaluation of $\protect\rho $, $p$, $\protect\omega $, $T$ and $%
f(Q,T)$:}

By using Eq. (\ref{eqn21}) in Eqs. (\ref{eqn26})-(\ref{eqn28}) the
expressions for $\rho $, $p$, $\omega $, $T$ and $f(Q,T)$ model are obtained
as follows:
\begin{widetext}
Energy density ($\rho $):
\begin{equation}
\begin{split}
\rho =& -\frac{1}{2(\gamma +4\pi )(\gamma +8\pi )}\left[ 36\beta \gamma
n(mt+nt_{0})^{2}t^{-4}t_{0}^{-2}+3\alpha \left( \gamma +8\pi \right) \left( 
\frac{m}{t_{0}}+\frac{n}{t}\right) ^{2}\right. \\
& +\left. 54\beta \left( \gamma +8\pi \right) \left( \frac{m}{t_{0}}+\frac{n%
}{t}\right) ^{4}+\alpha \gamma nt^{-2}\right] ,
\end{split}
\label{eqn29}
\end{equation}

Pressure ($p$):
\begin{equation}
\begin{split}
p=& -\frac{1}{2(\gamma +4\pi )(\gamma +8\pi )}\left[ \beta n\left( 108\gamma
+576\pi \right) \left( mt+nt_{0}\right) ^{2}t^{-4}t_{0}^{-2}-3\alpha \left(
\gamma +8\pi \right) \left( \frac{m}{t_{0}}+\frac{n}{t}\right) ^{2}\right. \\
& -\left. 54\beta \left( \gamma +8\pi \right) \left( \frac{m}{t_{0}}+\frac{n%
}{t}\right) ^{4}+\alpha n\left( 3\gamma +16\pi \right) t^{-2}\right] .
\end{split}
\label{eqn30}
\end{equation}
\end{widetext}

\begin{figure}[H]
\centering
\includegraphics[scale=0.67]{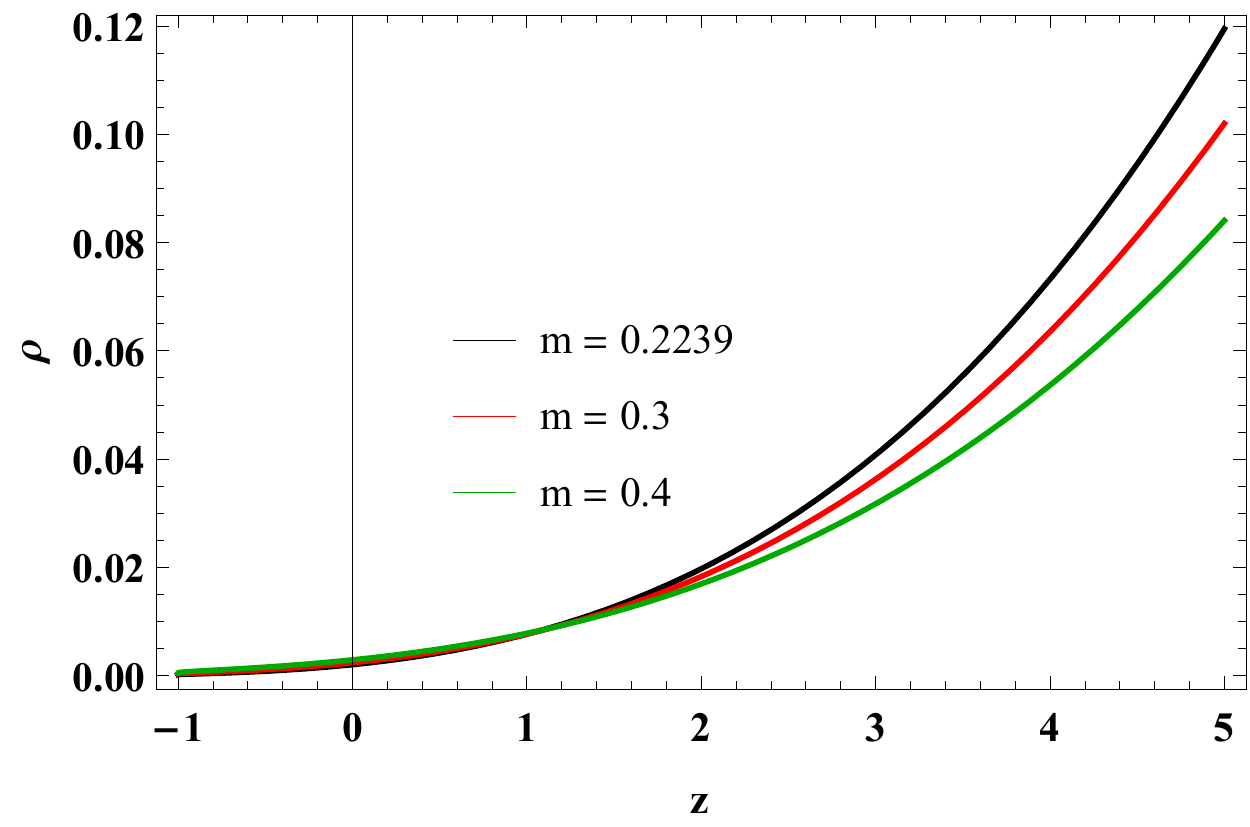}
\caption{Energy density $\left( \rho\right)  $ versus redshift $\left( z\right)  $ with $n=0.6886$, $\protect\alpha =-3$, $\protect\beta %
=0.003$ and $\protect\gamma =-\protect\pi $.}
\label{fig3}
\end{figure}

\begin{figure}[H]
\centering
\includegraphics[scale=0.67]{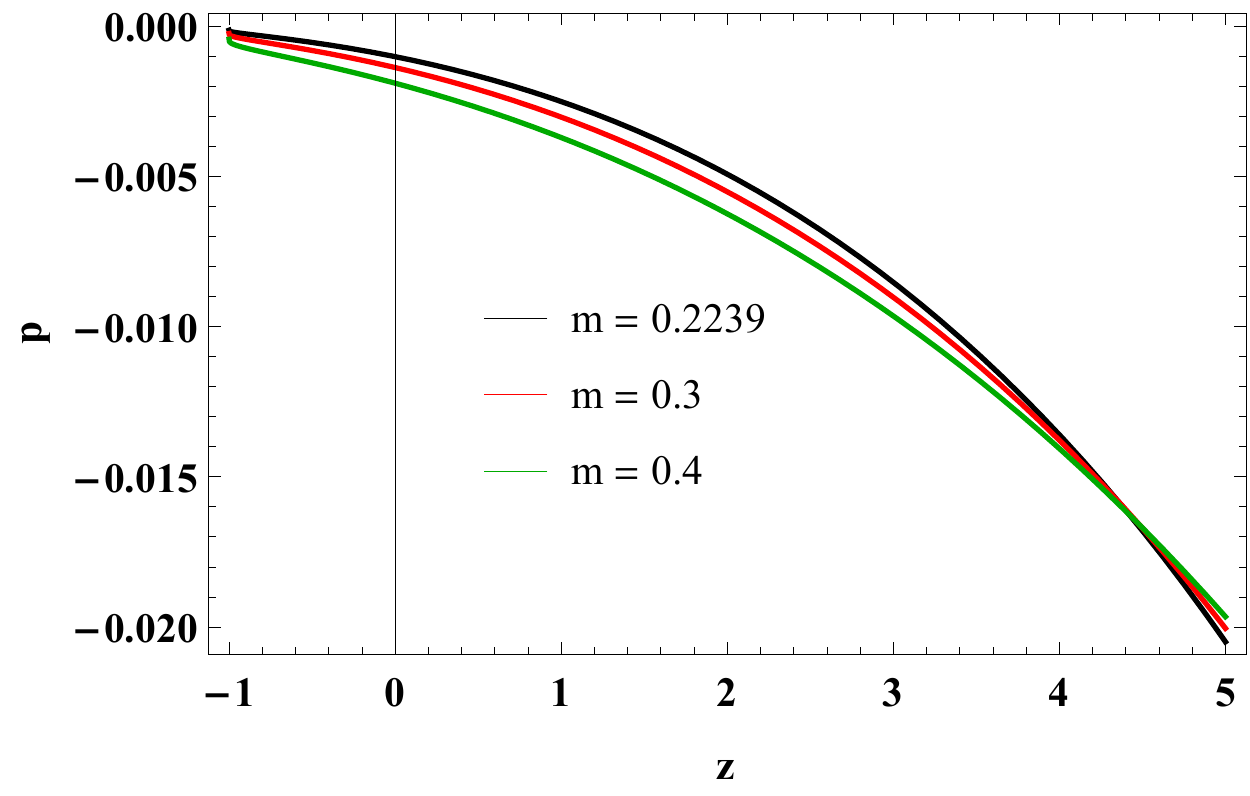}
\caption{Pressure $\left( p\right)  $ versus redshift $\left( z\right)  $  with $n=0.6886$, $\protect\alpha =-3$, $\protect\beta %
=0.003$ and $\protect\gamma =-\protect\pi $.}
\label{fig4}
\end{figure}

Fig. \ref{fig3} represents the evolution of the energy density of the
Universe as a function of redshift for three values of the parameter $%
m=0.2239,0.3,0.4$. From this figure, we can see that the energy density
remains positive for all $z$ values and is an increasing function of the
cosmological redshift. It starts with a positive value and approaches zero
when $z\rightarrow -1$. The pressure behavior as a function of redshift is
shown in Fig. \ref{fig4}, we observe that the pressure in the current model
is a decreasing function of the cosmological redshift, and it starts from a
large negative value and approaches zero at the present time. According to
recent observations, the Universe is in an accelerating expansion phase due
to the so-called dark energy that has negative pressure. Thus the pressure
for our model is consistent with recent observations.

\begin{figure}[H]
\centering
\includegraphics[scale=0.67]{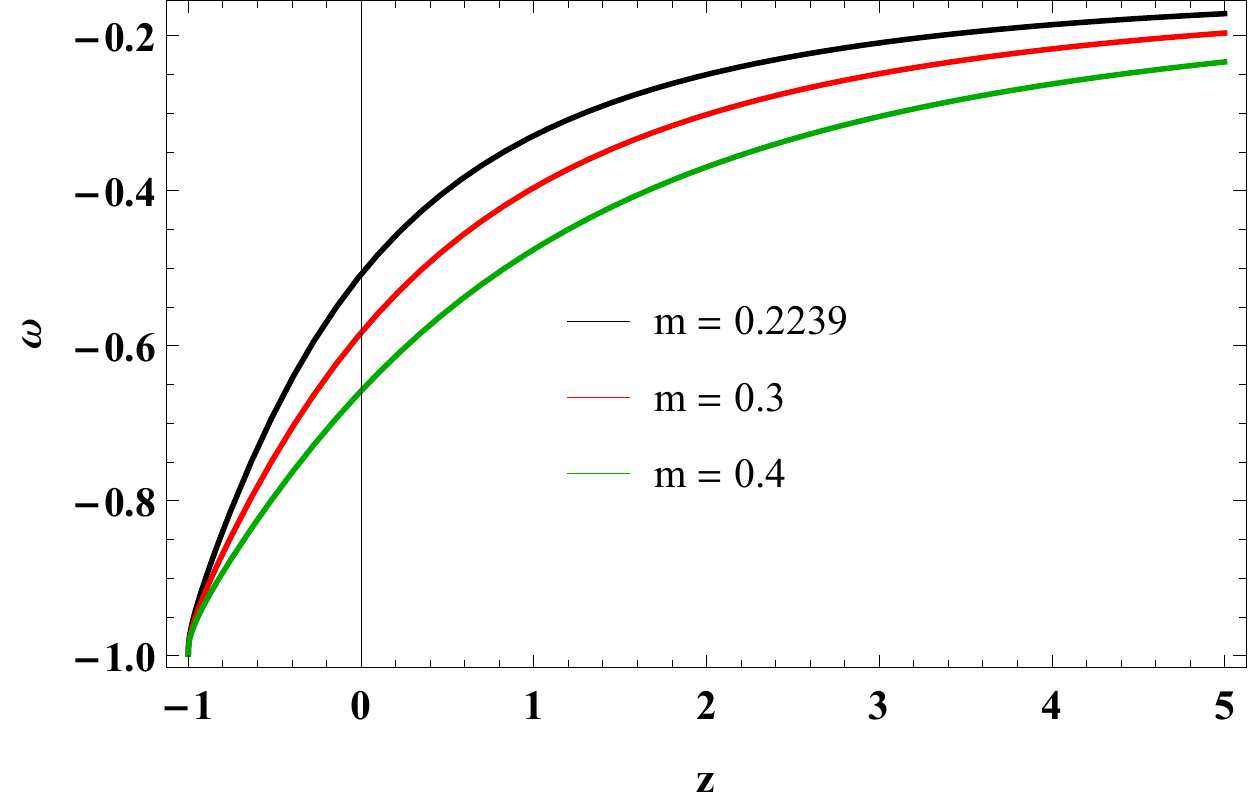}
\caption{EoS parameter $\left( \protect\omega \right) $ versus redshift $%
\left( z\right) $ with $n=0.6886$, $\protect\alpha =-3$, $\protect\beta %
=0.003$ and $\protect\gamma =-\protect\pi $.}
\label{fig5}
\end{figure}

\begin{widetext}
EoS parameter ($\omega $): 
\begin{equation}
\omega =\frac{\beta n\left( 108\gamma +576\pi \right) \left(
mt+nt_{0}\right) ^{2}t^{-4}t_{0}^{-2}-3\alpha \left( \gamma +8\pi \right)
\left( \frac{m}{t_{0}}+\frac{n}{t}\right) ^{2}-54\beta \left( \gamma +8\pi
\right) \left( \frac{m}{t_{0}}+\frac{n}{t}\right) ^{4}+\alpha n\left(
3\gamma +16\pi \right) t^{-2}}{36\beta \gamma n\left( mt+nt_{0}\right)
^{2}t^{-4}t_{0}^{-2}+3\alpha \left( \gamma +8\pi \right) \left( \frac{m}{%
t_{0}}+\frac{n}{t}\right) ^{2}+54\beta \left( \gamma +8\pi \right) \left( 
\frac{m}{t_{0}}+\frac{n}{t}\right) ^{4}+\alpha \gamma nt^{-2}}.
\label{eqn31}
\end{equation}
\end{widetext}

The EoS parameter is an essential tool for describing the epochs the
Universe has gone through, and understanding the nature of dark energy. This
parameter takes various values for each different model of dark energy. In
the case that the nature of dark energy is the cosmological constant ($%
\Lambda $CDM), $\omega =-1$. While if $-1<\omega <-0.33$, we say that the
nature of dark energy is quintessence, and $\omega <-1$, indicates phantom
nature of the model. From Fig. \ref{fig5}, we can see that $-1<\omega <-0.2$%
, it indicates quintessence nature of the model in the present, and in the
future the model approaches to the $\Lambda $CDM region. The current value
of $\omega $ is obtained as $\omega _{0}=-0.5079$ for the values of the
constants constrained by the observational Hubble parameter $H\left(
z\right) $ data i.e. $m=0.2239$ and $n=0.6886$. Thus, the current value of
EoS parameter of the model is consistent with Planck's 2018 results.

The value of the trace of energy-momentum tensor $T$ is obtained as

\begin{widetext}
\begin{equation}
\begin{split}
T=& 3p-\rho =\frac{1}{(\gamma +4\pi )(\gamma +8\pi )t^{4}t_{0}^{4}}\left[
2\left( 54\beta (\gamma +8\pi )m^{4}t^{4}+216\beta (\gamma +8\pi
)m^{3}nt^{3}t_{0}\right. \right. \\
& +\left. \left. 3m^{2}t^{2}t_{0}^{2}\left( \gamma \left( 12\beta
n(9n-2)+\alpha t^{2}\right) +8\pi \left( 18\beta n(6n-1)+\alpha t^{2}\right)
\right) +6mntt_{0}^{3}\left( \gamma \left( 12\beta n(3n-2)+\alpha
t^{2}\right) \right. \right. \right. \\
& +\left. \left. \left. 8\pi \left( 18\beta n(2n-1)+\alpha t^{2}\right)
\right) +nt_{0}^{4}\left( \gamma \left( 18\beta (3n-4)n^{2}+\alpha
(3n-2)t^{2}\right) \right. \right. \right. \\
& +\left. \left. \left. 12\pi \left( 36\beta (n-1)n^{2}+\alpha
(2n-1)t^{2}\right) \right) \right) \right]
\end{split}
\label{eqn32}
\end{equation}

Using the definition of non-metricity $Q$ for flat FLRW Universe and Eq. (%
\ref{eqn25}), the function $f\left( Q,T\right) $ is obtained as\newline

\begin{equation}
\begin{split}
f\left( Q,T\right) =& \frac{1}{(\gamma +4\pi )(\gamma +8\pi )t^{4}t_{0}^{4}}%
\left[ 4\left( 36\beta \left( \gamma ^{2}+9\pi \gamma +8\pi ^{2}\right)
m^{4}t^{4}+144\beta \left( \gamma ^{2}+9\pi \gamma +8\pi ^{2}\right)
m^{3}nt^{3}t_{0}\right. \right. \\
& +\left. \left. 3m^{2}t^{2}t_{0}^{2}\left( 16\pi ^{2}\left( 36\beta
n^{2}+\alpha t^{2}\right) +\gamma ^{2}\left( 12\beta n(6n-1)+\alpha
t^{2}\right) +2\pi \gamma \left( 36\beta n(9n-1)+5\alpha t^{2}\right)
\right) \right. \right. \\
& +\left. \left. 6mntt_{0}^{3}\left( 16\pi ^{2}\left( 12\beta n^{2}+\alpha
t^{2}\right) +\gamma ^{2}\left( 12\beta n(2n-1)+\alpha t^{2}\right) +2\pi
\gamma \left( 36\beta n(3n-1)+5\alpha t^{2}\right) \right) \right. \right. \\
& +\left. \left. nt_{0}^{4}\left( 36\beta \left( \gamma ^{2}+9\pi \gamma
+8\pi ^{2}\right) n^{3}-36\beta \gamma (\gamma +6\pi )n^{2}+3\alpha \left(
\gamma ^{2}+10\pi \gamma +16\pi ^{2}\right) nt^{2}-\alpha \gamma (\gamma
+6\pi )t^{2}\right) \right) \right]
\end{split}
\label{eqn33}
\end{equation}
\end{widetext}

\begin{figure}[ht]
\centering
\includegraphics[scale=0.67]{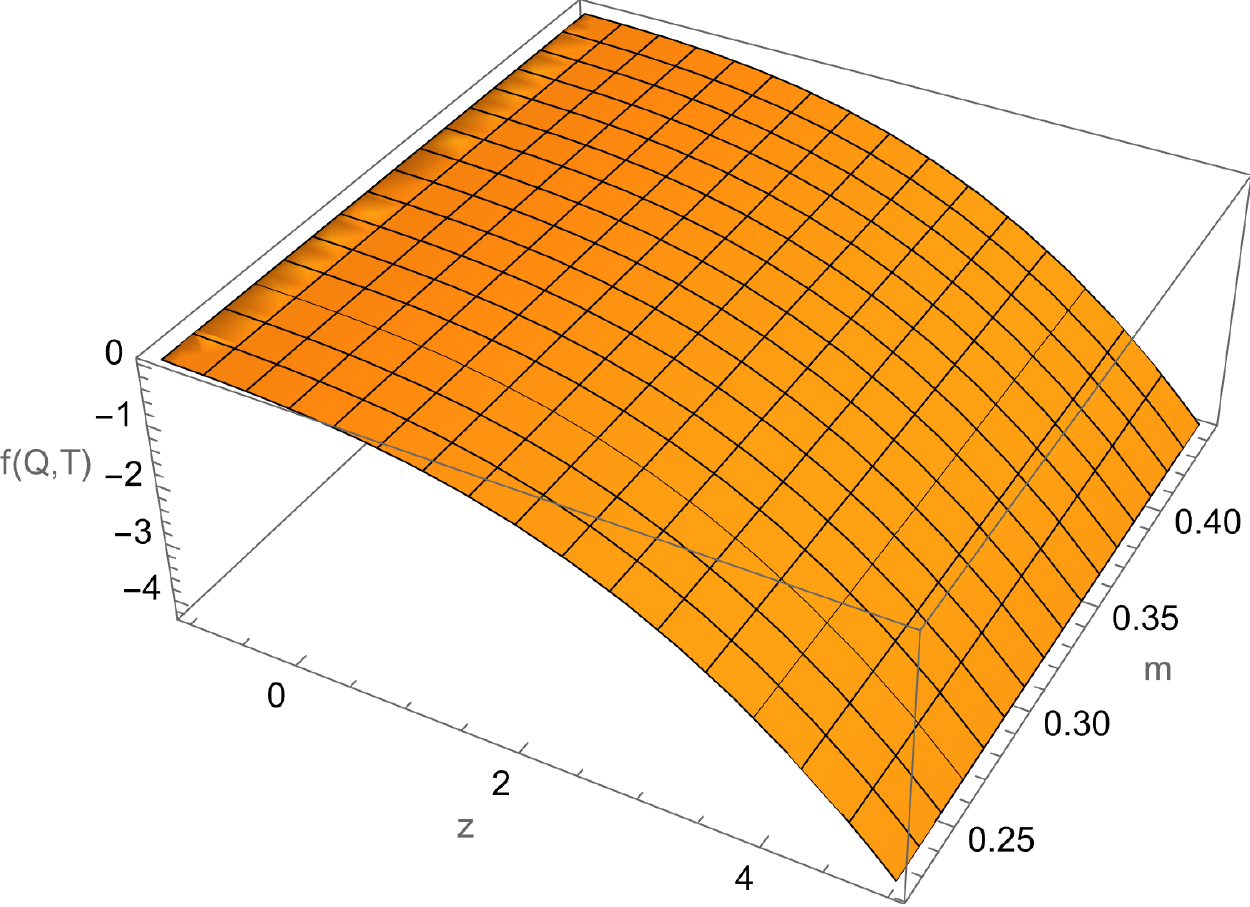}
\caption{$f(Q,T)$ gravity model versus versus redshift ($z$) and $m$.}
\label{fig6}
\end{figure}

The behavior of $f(Q,T)$ gravity model versus redshift ($z$) and $m$ is
clearly shown in Fig. \ref{fig6}.

\subsection{Energy conditions (ECs):}

We previously studied some cosmological parameters that plays an important
role in studying the evolution of the Universe, such as the deceleration
parameter, EoS parameter, etc. But in order to predict the cosmic
acceleration in modern cosmology, a set of energy conditions appeared that
relates the energy density of the Universe and pressure and can be derived
from equation of Raychaudhuri \cite{ref25}. In GR, the role of these energy
conditions is to prove the theorems for the existence of space-time
singularity and black holes \cite{ref26}. Several authors have worked on
energy conditions in various backgrounds \cite{ref20, ref21, ref22}. In this
paper, we will consider the famous energy conditions in order to check the
validity of the model in the context of cosmic acceleration. There are
different forms of energy conditions such as the weak energy conditions
(WEC), null energy condition (NEC), dominant energy conditions (DEC), and
strong energy conditions (SEC) are given for the content of the Universe in
form of a perfect fluid in $f\left( Q,T\right) $\ modified gravity\ as
follows

\begin{itemize}
\item WEC: $\rho \geq 0,$

\item NEC: $\rho +p\geq 0,$

\item DEC: $\rho -p\geq 0,$

\item SEC: $\rho +3p\geq 0.$
\end{itemize}

The significance of these energy conditions above shows that when the NEC is
violated, all other energy conditions are violated. This violation of the
NEC represents the depletion of energy density as the Universe expands.
Also, the violation of the SEC represents the acceleration of the Universe.
We can see this from the standard Friedmann equations, in order to explain
the late-time cosmic acceleration with $\omega \simeq -1$, it must be $\rho
+3p=\rho \left( 1+3\omega \right) <0$. In Fig. \ref{fig7}, we can see the
evolution of the energy conditions WEC, NEC, DEC and SEC as functions of the
cosmological redshift and $m$, respectively. From the figures, we observe
that WEC, NEC and DEC are satisfied while the SEC is violated in the present
and future. Hence, the violation of SEC leads to the acceleration of the
Universe (see Fig. \ref{fig7})

\begin{figure*}[ht]
\begin{minipage}{0.45\linewidth}
  \centerline{\includegraphics[scale=0.65]{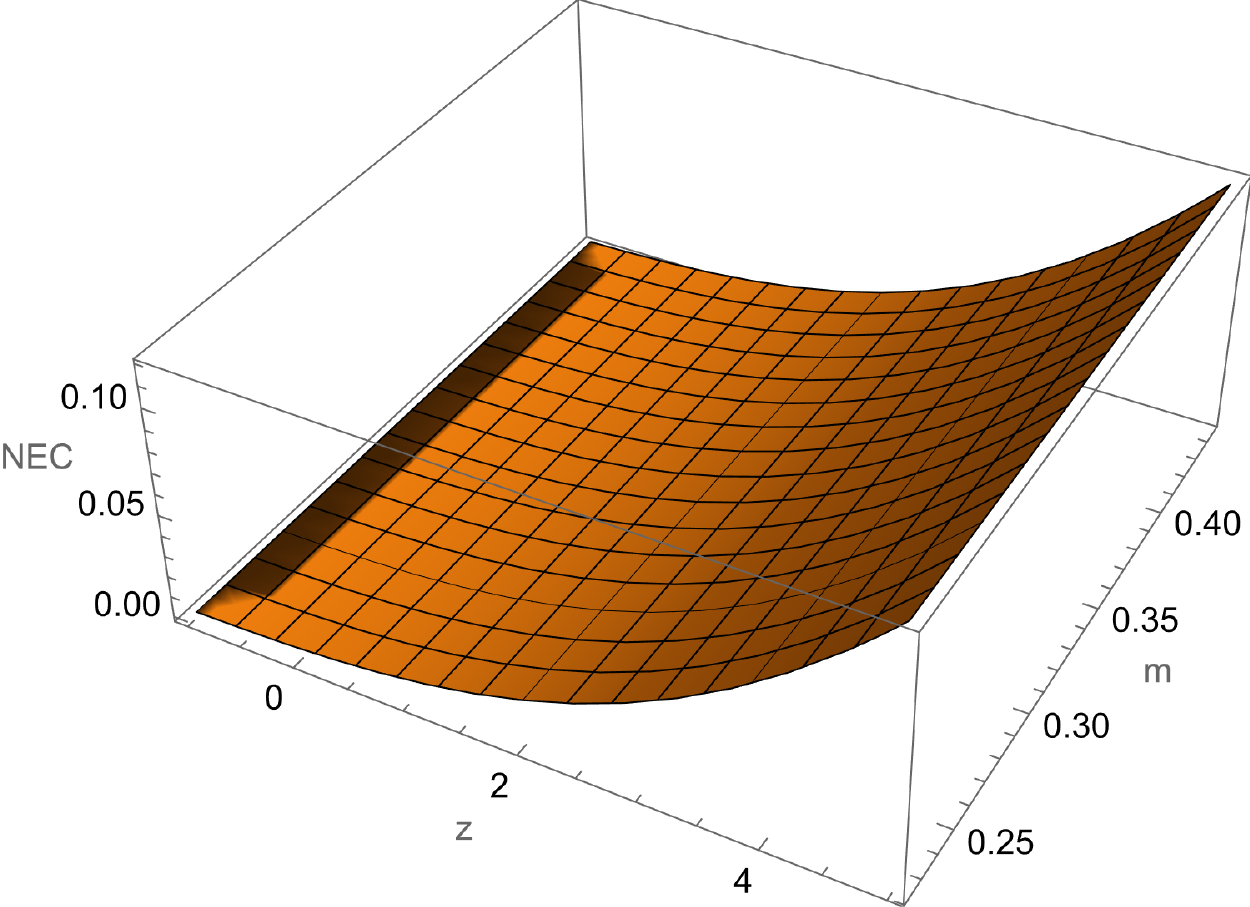}}
  \centering{ $(\rho +p\geq 0$)}
 \end{minipage}\hfill 
\begin{minipage}{0.45\linewidth}
  \centerline{\includegraphics[scale=0.65]{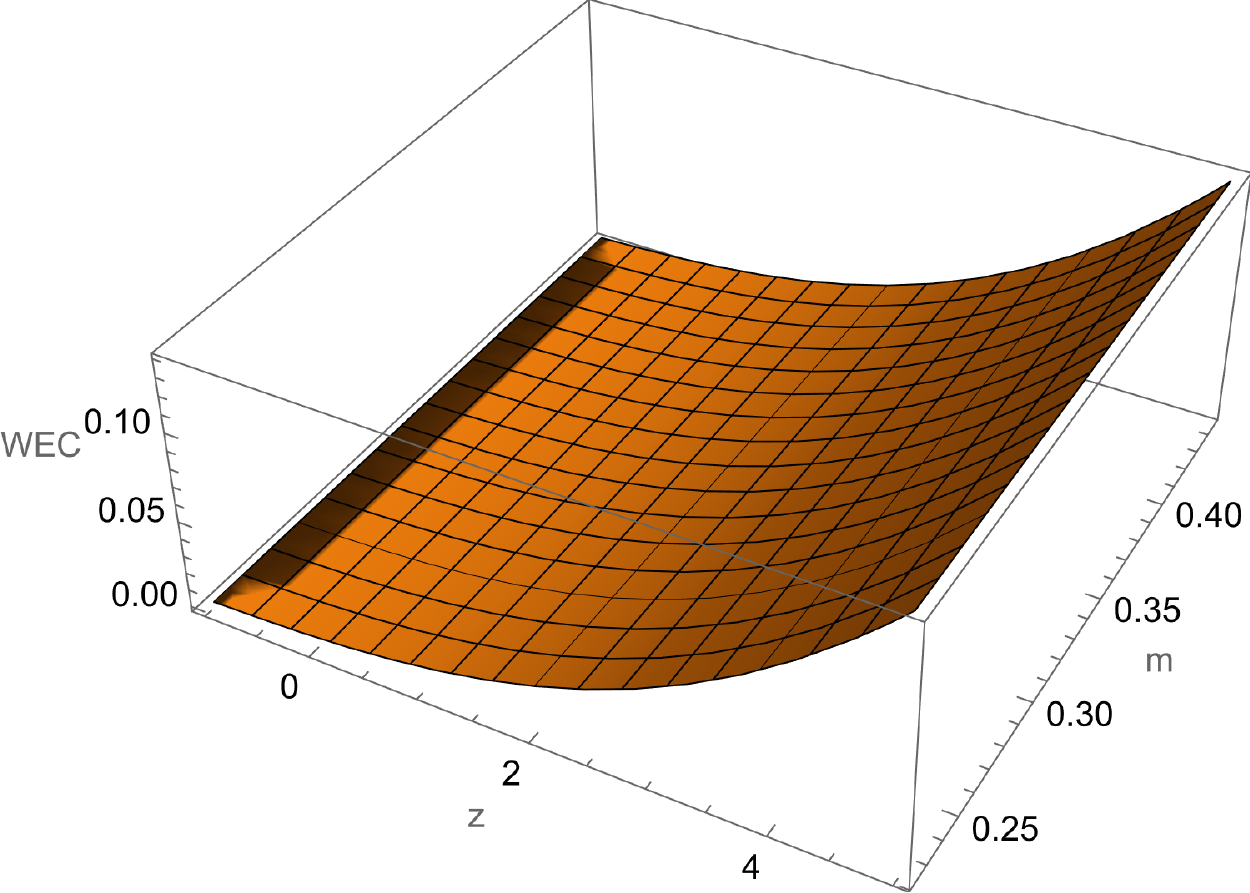}}
  \centering{ ($\rho \geq 0$)}
 \end{minipage}
\end{figure*}

\begin{figure*}[ht]
\begin{minipage}{0.45\linewidth}
  \centerline{\includegraphics[scale=0.65]{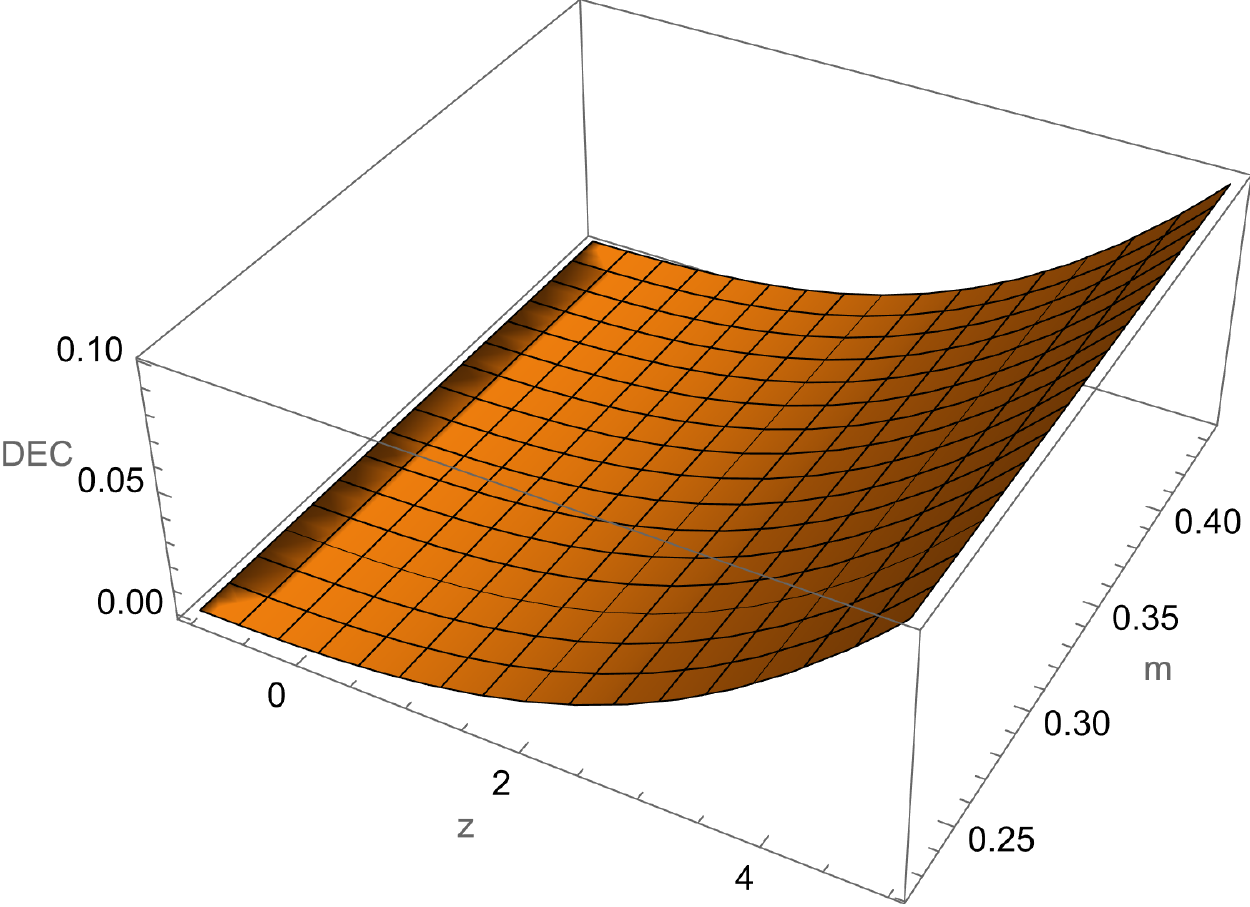}}
  \centering{( $\rho -p\geq 0$)}
 \end{minipage}\hfill 
\begin{minipage}{0.45\linewidth}
  \centerline{\includegraphics[scale=0.65]{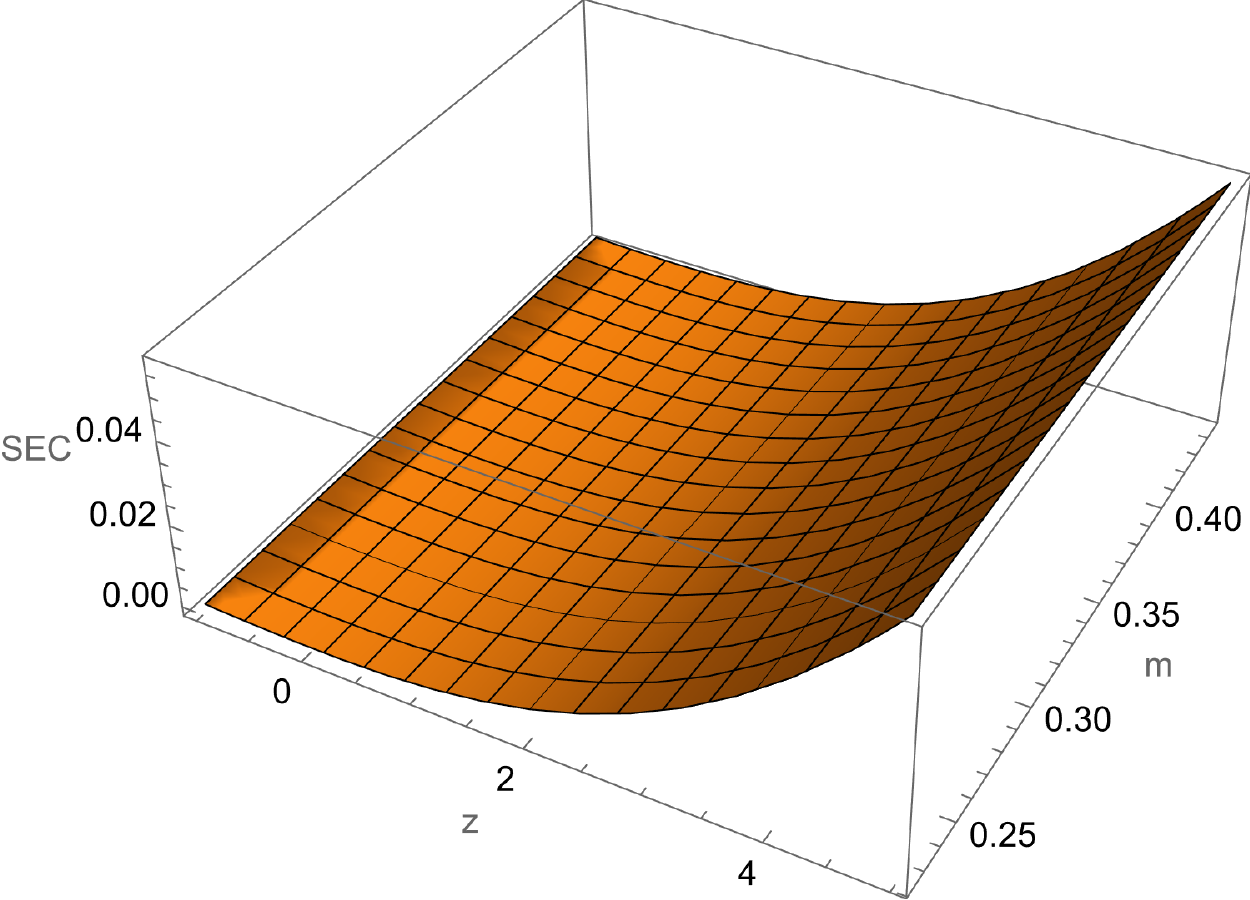}}
  \centering{($\rho +3p\geq 0$)}
 \end{minipage}
\caption{Energy conditions versus redshift $\left( z\right) $.}
\label{fig7}
\end{figure*}

\subsection{Cosmographic jerk parameter}

The jerk parameter is one of the basic physical quantities to explain the
dynamics of the Universe. The Jerk parameter is a dimensionless third
derivative of the scale factor $a\left( t\right) $ with respect to cosmic
time $t$ and is specified as \cite{ref22}

\begin{equation}
j=\frac{\overset{...}{a}}{aH^{3}}.  \label{eqn34}
\end{equation}

Eq. (34) can be written in terms of a deceleration parameter $q$ as

\begin{equation}
j=q+2q^{2}-\frac{\overset{.}{q}}{H}.  \label{eqn35}
\end{equation}

Using Eqs. (\ref{eqn21}) and (\ref{eqn22}), the jerk parameter for our model
is

\begin{widetext}
\begin{equation}
j=\left[ m^{3}t^{3}+3m^{2}nt^{2}t_{0}+3m(n-1)ntt_{0}^{2}+n\left(
n^{2}-3n+2\right) t_{0}^{3}\right] (mt+nt_{0})^{-3}  \label{eqn36}
\end{equation}
\end{widetext}

The value of the jerk parameter is $j=1$ for $\Lambda CDM$ model. The
Universe is transitioning from an early deceleration phase to the current
phase of acceleration with a positive jerk parameter $j_{0}>0$ and a
negative DP $q_{0}<0$ corresponding to $\Lambda CDM$ model. From Fig. \ref%
{fig8} we can see that the jerk parameter remains positive for $m=0.2239$
and $n=0.6886$ and approaches $1$ later. The current jerk parameter value $%
j_{0}$ is positive. Thus, our model is similar to the $\Lambda CDM$ model in
the future.

\begin{figure}[H]
\centerline{\includegraphics[scale=0.67]{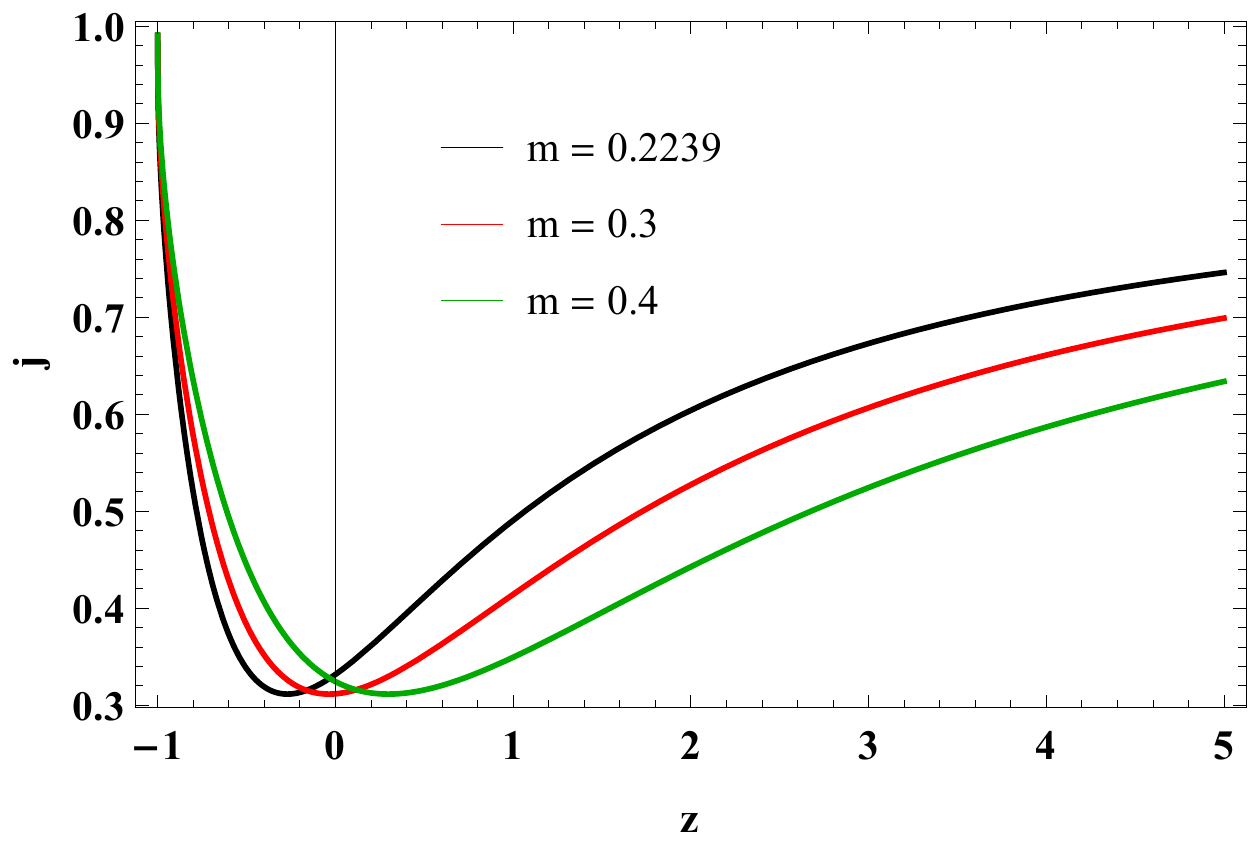}}
\caption{Jerk parameter ($j$) versus redshift ($z$).}
\label{fig8}
\end{figure}

\section{Conclusion}

\label{sec6}

In the current analysis, we studied one of interesting extension of $f(Q)$
gravity theory in form $f(Q,T)$ for geometric alternatives to dark energy in
which the term $Q$ is non-metricity scalar and $T$ is the trace of the
matter energy-momentum tensor using a source as perfect fluid. In this
analysis, we found the best fit value of the model parameters $m$ and $n$ or
constrain this parameters using some observational datasets such as $57$
data points in the cosmological redshift range $0.07\leq z\leq 2.42$, $31$
points from the differential age method (DA method)\ and the other $26$
points were evaluated using BAO data and other methods. In the derived
model, the transition of the Universe from the deceleration phase $\left(
q>0\right) $ to the acceleration phase $\left( q<0\right) $ with the
transition redshift $z_{tr}=05234$ for $m=0.2239$ which is the evidence that
the transition redshift value for our model is in conformity with the
observational data.

In addition, we have studied the evolution of the energy density of the
Universe as a function of redshift which remains positive for all $z$ and is
an increasing function of the cosmological redshift. The pressure behavior
as a function of redshift of the model was also studied and is a decreasing
function of the cosmological redshift, and it starts from a large negative
value and approaches zero at the present time. Thus, the model is consistent
with recent observations. As for the EoS parameter behavior of our model, it
is in the range $-1<\omega <-0.2$ which indicates a quintessence nature of
the model at present, and in the future the model approaches $\Lambda CDM$
region whereas the current value of $\omega $ is obtained as $\omega
_{0}=-0.5079$ for the values of the constants $m=0.2239$ and $n=0.6886$.
Therefore, the current value of EoS parameter of the model is consistent
with Planck's 2018 results. Finally, the energy conditions: WEC, NEC, and
DEC are satisfied while the SEC is violated in the present and future. The
violation of SEC leads to the acceleration of the Universe while the jerk
parameter remains positive and approaches $1$ later. Thus, our model is
similar to the $\Lambda CDM$ model in the future.

\section*{Acknowledgments}

We are very much grateful to the honorary referee and the editor for the
illuminating suggestions that have significantly improved our work in terms
of research quality and presentation. This research was funded by the
Science Committee of the Ministry of Education and Science of the Republic
of Kazakhstan (Grant No. AP09058240). \newline

\textbf{Data availability} There are no new data associated with this article%
\newline

\textbf{Declaration of competing interest} The authors declare that they
have no known competing financial interests or personal relationships that
could have appeared to influence the work reported in this paper.\newline

\begin{table}[tbp]
\begin{center}
\begin{tabular}{|c|c|c|c|c|c|}
\hline\hline
$z$ & $H(z)$ & $\sigma _{H}$ & $z$ & $H\left( z\right) $ & $\sigma _{H}$ \\ 
\hline\hline
0.070 & 69 & 19.6 & 0.4783 & 80 & 99 \\ \hline
0.90 & 69 & 12 & 0.480 & 97 & 62 \\ \hline
0.120 & 68.6 & 26.2 & 0.593 & 104 & 13 \\ \hline
0.170 & 83 & 8 & 0.6797 & 92 & 8 \\ \hline
0.1791 & 75 & 4 & 0.7812 & 105 & 12 \\ \hline
0.1993 & 75 & 5 & 0.8754 & 125 & 17 \\ \hline
0.200 & 72.9 & 29.6 & 0.880 & 90 & 40 \\ \hline
0.270 & 77 & 14 & 0.900 & 117 & 23 \\ \hline
0.280 & 88.8 & 36.6 & 1.037 & 154 & 20 \\ \hline
0.3519 & 83 & 14 & 1.300 & 168 & 17 \\ \hline
0.3802 & 83 & 13.5 & 1.363 & 160 & 33.6 \\ \hline
0.400 & 95 & 17 & 1.430 & 177 & 18 \\ \hline
0.4004 & 77 & 10.2 & 1.530 & 140 & 14 \\ \hline
0.4247 & 87.1 & 11.2 & 1.750 & 202 & 40 \\ \hline
0.4497 & 92.8 & 12.9 & 1.965 & 186.5 & 50.4 \\ \hline
0.470 & 89 & 34 &  &  &  \\ \hline\hline\hline
$z$ & $H\left( z\right) $ & $\sigma _{H}$ & $z$ & $H\left( z\right) $ & $%
\sigma _{H}$ \\ \hline\hline
0.24 & 79.69 & 2.99 & 0.52 & 94.35 & 2.64 \\ \hline
0.30 & 81.7 & 6.22 & 0.56 & 93.34 & 2.3 \\ \hline
0.31 & 78.18 & 4.74 & 0.57 & 87.6 & 7.8 \\ \hline
0.34 & 83.8 & 3.66 & 0.57 & 96.8 & 3.4 \\ \hline
0.35 & 82.7 & 9.1 & 0.59 & 98.48 & 3.18 \\ \hline
0.36 & 79.94 & 3.38 & 0.60 & 87.9 & 6.1 \\ \hline
0.38 & 81.5 & 1.9 & 0.61 & 97.3 & 2.1 \\ \hline
0.40 & 82.04 & 2.03 & 0.64 & 98.82 & 2.98 \\ \hline
0.43 & 86.45 & 3.97 & 0.73 & 97.3 & 7.0 \\ \hline
0.44 & 82.6 & 7.8 & 2.30 & 224 & 8.6 \\ \hline
0.44 & 84.81 & 1.83 & 2.33 & 224 & 8 \\ \hline
0.48 & 87.90 & 2.03 & 2.34 & 222 & 8.5 \\ \hline
0.51 & 90.4 & 1.9 & 2.36 & 226 & 9.3 \\ \hline
\end{tabular}%
\end{center}
\caption{57 points of $H(z)$ data: $31$ (DA) and $26$ (BAO+other) 
\protect\cite{ref24}.}
\label{tab1}
\end{table}

\end{document}